\documentclass[aps,pra,preprint,groupeaddress,showpacs]{revtex4}
\usepackage{graphicx}
\usepackage{dcolumn}
\usepackage{longtable}
\begin{document}
\preprint{Version 3.3}
\title{Recoil corrections in the hydrogen isoelectronic sequence}
\author{G.S. Adkins}
\email[]{gadkins@fandm.edu}
\affiliation{Department of Physics, Franklin and Marshall College,
Lancaster, PA 17604}

\author{J. Sapirstein}
\email[]{jsapirst@nd.edu}
\affiliation{Department of Physics, University of Notre Dame, Notre Dame, IN 46556}

\date{\today}
\begin{abstract}
A version of the Bethe-Salpeter equation appropriate for calculating recoil
corrections in highly charged
hydrogenlike ions is presented. The nucleus is treated as a scalar particle of charge $Z$,
and the electron treated relativistically. The known recoil corrections of order
${m^2 \over M} (Z\alpha)^4$ are derived in both this formalism and in NRQED.
\end{abstract}

\maketitle

\section{introduction}

Precision tests of QED in atoms were first carried out for hydrogen, and have been extended
to other one-electron systems such as positronium and muonium, as well as helium \cite
{Drakebook}, lithium \cite{lithium}, and even beryllium \cite{beryllium}. In all these cases 
the nuclear charge $Z$ is low, so the basic expansion parameter of bound state QED, $Z\alpha$, 
is a small quantity. For this reason techniques in which the smallness of this parameter is 
exploited have been refined in sophistication over the years, culminating in the present 
widespread use of effective field theories such as NRQED \cite{NRQED} and the
effective Hamiltonian method \cite{Pach1}. However, at the same time experiments of increasing 
precision have been carried out on both highly charged hydrogenlike ions and also ions with 
more electrons, where as an example of the accuracy achieved at the highest $Z$ we note the 
recent determination \cite{Beiersdorfer} of the $2p_{1/2}-2s_{1/2}$ transition energy in lithiumlike 
uranium, 
\begin{equation}
E_{2p_{1/2}}-E_{2s_{1/2}} = 280.645(15)~{\rm eV}.
\end{equation}
As the expansion parameter $Z\alpha$ is no longer small in this case, techniques in which
an expansion in it is avoided are necessary. In the non-recoil limit, in which
the nuclear mass is taken to infinity, Furry representation QED \cite{Furry} allows a systematic 
Feynman diagram based treatment of highly charged ions. A central structure in
this approach is the electron propagator in a Coulomb field, the Dirac-Coulomb propagator,
which satisfies the equation
\begin{equation}
\left [ \left (E + {Z \alpha \over {\left \vert \vec x \right \vert}} \right ) \gamma_0 + i \vec \gamma \cdot \vec \nabla - m \right ] 
S_F(\vec x, \vec y; E) = \delta^3 \left ( \vec x- \vec y \right ).
\label{DiracCoulombprop}
\end{equation}
As first pointed out by Wichmann and Kroll \cite{W-K} for the vacuum polarization, and by
Brown, Langer and Schaefer \cite{BLS} for the self energy, treating this propagator exactly 
using numerical methods allows a determination of the Lamb shift that automatically accounts for all
orders of an expansion in $Z\alpha$. 

When applied to lithiumlike uranium, use of this propagator gives a one-loop Lamb shift contribution
(including screening corrections) to the $2p_{1/2}-2s_{1/2}$ splitting of -41.793 eV, which when 
combined with the nonradiative energy shift of 322.231 eV leaves a 0.207 eV discrepancy with 
experiment. This can be used
to infer the two-loop Lamb shift, which has recently been calculated for the ground state
of hydrogenic ions \cite{2loop}, but before this can be done recoil terms, the subject we wish 
to address in this paper, must be reliably calculated.

Recoil effects have of course been treated for low $Z$ atoms, but again the techniques
cannot be directly extended to high $Z$ ions. The general level of treatment of these
small corrections in this latter case is to scale the overall energies by a factor of $\mu/m$, 
where the reduced mass $\mu$ is defined in the usual way in terms of the electron mass $m$ and the 
nuclear mass $M$,
\begin{equation}
\mu = {m M \over m+M}.
\end{equation}
This has a relatively small effect for the transition we are discussing, amounting to
only -0.006 eV, below the experimental error. A larger effect comes from the mass-polarization 
operator,
\begin{equation}
H_{MP} = \sum_{i<j} { \vec p_i \cdot \vec p_j \over M}.
\end{equation}
When this term is evaluated in a realistic potential it contributes -0.081 eV,
for a total lowest order effect of -0.087 eV, 42 percent of the discrepancy. To
accurately infer the two-loop Lamb shift, a more sophisticated treatment of recoil effects is clearly
needed. The only such treatment we are aware of is that given by Shabaev \cite{Victor} and
collaborators \cite{Victors}, who in fact find significant corrections to the above
result. The present paper is intended to lay the 
groundwork for an alternative approach. We will restrict our attention to the hydrogen 
isoelectronic sequence, and in addition restrict our attention to diagrams that contribute 
in the low $Z$ case to order $m^2(Z\alpha)^4/M$, leaving the treatment of a more complete  
set of diagrams, along with the treatment of the many-electron problem, for a later paper.

While considerable progress has been made in QED in recent years with the use of effective field
theories, these rely on expanding around the nonrelativistic Schr\"{o}dinger equation, which, as just
discussed, is not appropriate for highly charged ions. However, the Bethe-Salpeter formalism
\cite{B-S}, introduced first to treat the binding of the deuteron and shortly afterwards
applied to the atomic problem by Salpeter \cite{Salpeter}, allows the problem to be treated
in a systematic manner. However, this equation is famously difficult to apply, and most applications
rely on expanding around the nonrelativistic limit, which we wish to avoid.

The treatment given by Shabaev is fairly complicated, and we wish to provide a cross check by
introducing as simple a formalism as possible. This can be done
by slightly modifying a formalism introduced by Lepage \cite{Lepage}, and it is 
this approach we will now describe.

The plan of the paper is to set up in the following section a three-dimensional formalism equivalent 
in rigor to the Bethe-Salpeter equation. In the next section the one and two photon exchange diagrams
that contribute to order $m^2(Z\alpha)^4/M$ will be evaluated in Coulomb gauge, and their 
nonrelativistic limit will be taken. This will be followed by a NRQED treatment, and in the 
conclusion we will describe how a calculation relevant to highly charged ions can be carried out. 

\section{Formalism}

It has been known for quite some time \cite{Fronsdal} that there is an arbitrary number of bound 
state equations equivalent in rigor to the original form of the Bethe-Salpeter equation, but that 
are effectively three-dimensional. Many practical calculations have used formulations that also 
incorporate the Schr\"{o}dinger equation \cite{Caswell-Lepage}, \cite{Adkins-Fell}.  However, for 
the problem we are considering a relativistic approach is demanded.  We note that a fairly 
detailed discussion of a number of notational and formal issues involved with the use of three-dimensional forms of the Bethe-Salpeter equation is given in Ref. \cite{Adkins-Fell}, to which we 
refer the reader interested in more detail.

The spectrum of many high-$Z$ ions has been studied, and it would be impractical
to consider the differing spins of each nucleus. For this reason we simply treat the nucleus as
a spinless particle of charge $Z|e|$. The Feynman rules for the electrodynamics of a spin-0 particle
involve the coupling $iZ|e|(p+p')_{\mu}$ for the one-photon vertex, and $2i(Ze)^2 g_{\mu \nu}$ for
the seagull vertex. The $\mu =0 $ component of the one-photon vertex for a nucleus close to mass
shell will then be dominated by the factor $2M$, where $M$ is the mass of the nucleus. Hyperfine
effects associated with nuclear spin can be treated separately. We can also model the finite size
of the nucleus by replacing the nuclear charge $Z$ with a form factor $Z({\vec q\,}^2)$ if desired, but
this will not be done in this paper.

We now consider the truncated two-particle Green's function for the scattering of an electron and
nucleus. We define the  initial and final electron three-momenta as $\vec k$ and $\vec l$ 
respectively, and work in the center of mass so that the corresponding nuclear momenta 
are $-\vec k$ and $-\vec l$. For the fourth component of momentum we choose $E_1 + k_0$ 
and $E_1 + l_0$  for the
electron line and $E_2 -k_0$ and $E_2-l_0$ for the nuclear line, where $E_1$ and $E_2$ will
be chosen close to the electron and nuclear masses, and when the total center of mass energy
$E = E_1 + E_2$ is a bound state energy a pole will be present. The formalism to be
described below in its simplest form leads to a perturbation expansion about 
$E_1 = \epsilon$ and $E_2 = M$, with $\epsilon$ equal to the Dirac bound state energy,
\begin{eqnarray}
\label{reference_energies}
{\epsilon} &=& m \Biggl [ 1+ \Biggl ( \frac{Z \alpha}{n-(j+1/2)+\sqrt{(j+1/2)^2-(Z \alpha)^2}} \Biggr )^2 \Biggr ]^{-1/2} \cr
&=& m \Biggl \{ 1- \frac{(Z \alpha)^2}{2n^2} - \frac{(Z \alpha)^4}{2 n^3} \Bigl ( \frac{1}{j+1/2} - \frac{3}{4n} \Bigr ) + O((Z \alpha)^6) \Biggr \} \, ,
\end{eqnarray}
giving a total bound state energy  $E = M+\epsilon = M + m - m\frac{(Z \alpha)^2}{2n^2} + ...$
However, since this does not incorporate the
known reduced mass dependence of the nonrelativistic binding energy, we will instead arrange
the formalism so that $E_1 = {\cal E} = {\mu \over m} \epsilon$ and 
$E_2 = M + m - \mu \equiv \tilde{M}$. We incorporate these energies into the 
four-vectors $P_1 = ( {\cal E}, \vec 0 \,)$ and
$P_2 = ( \tilde{M}, \vec 0 \,)$. If we also define four vectors $k = (k_0, \vec k \,)$ and
$l = (l_0, \vec l \,)$, the initial electron and nuclear momenta are $P_1 + k$, $P_2 -k$,
and the final momenta $P_1 + l$, $P_2 - l$.
The truncated two particle Green's function obeys the equation
\begin{eqnarray}
G_T(P_1+k, P_2-k; P_1+l,P_2-l) = i K(P_1+k,P_2-k;P_1+l,P_2-l) + \nonumber \\
\int {d^4 q \over (2\pi)^4}  K(P_1+k,P_2-k;P_1+q,P_2-q) S(q) G_T(P_1+q,P_2-q;P_1+l,P_2-l),
\end{eqnarray}
where $K$ represents all two-particle irreducible kernels and the spin-1/2--spin-0 
two-particle propagator has the form
\begin{equation}
S(q) = \frac{i}{\gamma(P_1+q)-m+i \epsilon}  \times \frac{i}{(P_2-q)^2-M^2+i \epsilon} \, .
\end{equation}
For brevity in the following we will write the above as
\begin{equation}
G_T(k,l;E) = i K(k,l;E) + \int {d^4 q \over (2\pi)^4} K(k,q;E) S(q) G_T(q,l;E).
\end{equation}

The main point of all simplifications of the Bethe-Salpeter formalism is that
we can replace the relatively complicated two-particle propagator $S$ with a simplified form $S_0$
and write
\begin{equation}
G_T(k,l;E) = i \bar{K}(k,l;E) + \int {d^4 q \over (2\pi)^4} 
\bar{K}(k,q;E) S_0(q) G_T(q,l;E),
\label{eqn5}
\end{equation}
which serves to define $\bar{K}$ through
\begin{eqnarray}
\bar{K}(k,l;E) = K(k,l;E) + \int{d^4 q \over (2\pi)^4} K(k,q;E) (S(q)-S_0(q))
K(q,l;E) + \nonumber \\ \int {d^4 q \over (2\pi)^4} \int {d^4 p \over (2\pi)^4}
K(k,q;E)(S(q)-S_0(q))K(q,p;E)(S(p)-S_0(p))K(p,l;E) + ...
\label{Kbardef}
\end{eqnarray}
Our choice for $S_0(q)$ is
\begin{eqnarray}
S_0(q) & = & { \pi \delta(q_0) \over \tilde{M}}  \frac{i}{{\cal E} \gamma_0 - \vec \gamma \cdot \vec q
-\mu+i \epsilon} \nonumber \\
& \equiv & 
{i \pi \delta(q_0) \over \tilde{M}} S_0(\vec q).
\end{eqnarray}
As mentioned above, we could have chosen another form with ${\cal E}$ and $\mu$ replaced
with $\epsilon$ and $m$, which would lead to the Dirac equation with mass $m$ in the $M \rightarrow \infty$ limit.   Our method will lead to a Dirac equation with reduced mass $\mu$ in that limit.  
We note that this method of building in the reduced mass is relatively simple, in particular
requiring no rescaling of coupling constants. The delta function 
we have chosen differs from that of Ref. \cite{Lepage}, in which a delta function that puts the 
nucleus on mass shell is chosen. While the latter choice has a number of advantages when Feynman 
gauge is used, we use Coulomb gauge in this calculation, and putting the nucleus on-shell is not 
needed. At this point we can go to a completely three-dimensional formalism by choosing 
$k_0 = l_0 = 0$, a choice that has no effect on the location of the bound state poles, and which 
allows us to replace the four vectors $k$ and $l$ with $\vec k$ and $\vec l$. In this case 
Eq.~(\ref{eqn5}) takes the three-dimensional form
\begin{equation}
G_T(\vec k,\vec l;E) = i \bar{K}(\vec k, \vec l;E) + {1 \over 2 \tilde{M}} \int {d^3 q \over (2\pi)^3} 
i \bar{K}(\vec k, \vec q;E) S_0(\vec q\, ) G_T(\vec q,\vec l;E).
\end{equation}

One gets to a bound state equation by creating an untruncated Green's 
function $\bar{G}(\vec k, \vec l;E)$ defined
through
\begin{equation}
\bar{G}(\vec k, \vec l; E) ={ 1 \over 2 \tilde{M}} S_0(\vec k \,) (2\pi)^3  {\delta}^3(\vec k - \vec l \, ) + 
{1 \over 4 \tilde{M}^2} S_0(\vec k \, ) G_T(\vec k, \vec l;E) S_0(\vec l \, )
\end{equation}
that satisfies
\begin{equation}
\bar{G}(\vec k, \vec l; E) = {1 \over 2 \tilde{M}} S_0(\vec k \, ) (2\pi)^3 {\delta}^3(\vec k - \vec l \, ) + 
S_0(\vec k \, ){1 \over 2 \tilde{M}} \int {d^3 q \over (2\pi)^3} i\bar{K}(\vec k, \vec q;E) 
\bar{G}(\vec q, \vec l;E).
\end{equation}
While this function differs from the Bethe-Salpeter untruncated Green's function, it has poles at
exactly the same total energy \cite{comment1}. In order to obtain a solvable problem, we
now introduce the simpler equation
\begin{equation}
G_0(\vec k, \vec l; E) = {1 \over 2 \tilde{M}} S_0(\vec k \, ) (2\pi)^3 {\delta}^3(\vec k - \vec l \, ) + 
S_0(\vec k \, ){1 \over 2 \tilde{M}} \int {d^3 q \over (2\pi)^3} iK_{1C}(\vec k, \vec q;E) 
G_0(\vec q, \vec l;E),
\end{equation}
which can be written, because the kernel $K_{1C}$ for one Coulomb photon exchange is
\begin{equation}
K_{1C} = { 4 \pi i Z \alpha \over {| \vec k - \vec q \, |^2}} 2 \tilde{M} \gamma_0,
\end{equation}
as
\begin{equation}
({\cal E} \gamma_0 - \vec \gamma \cdot \vec k - \mu) G_0(\vec k, \vec l; E) = 
{1 \over 2 \tilde{M}}(2\pi)^3 {\delta}^3(\vec k - \vec l \, ) - 4 \pi Z \alpha
\int \frac{d^3 q}{(2 \pi)^3}  {1 \over |\vec k - \vec q \, |^2} \gamma^0 G_0(\vec q, \vec l;E),
\label{basic}
\end{equation}
where we have multiplied by $S^{-1}_0(\vec k \, )$. 

In the following section we will discuss the effect of expanding $\bar{G}$ about $G_0$,
but here restrict our attention to the latter function, which, except for the factor
of ${1 \over 2 \tilde{M}}$ multiplying the delta function, is precisely the momentum space form
of Eq.~(\ref{DiracCoulombprop}) with the electron mass replaced with the reduced mass.
It has the spectral representation
\begin{equation}
G_0(\vec k, \vec l; E) = \sum_n {\psi_n(\vec k \, ) \bar{\psi}_n(\vec l \, ) 
\over {{\cal E} - {\cal E}_n}},
\end{equation}
where $\psi_n(\vec l)$ is the solution to the Dirac equation with the usual normalization
factor multiplied by $\sqrt{1 \over 2 \tilde{M}}$, and has poles when $E  = \tilde{M}+{\cal E}_n
\equiv E_0$.
To illustrate, the ground state (g) wave function has energy 
\begin{equation}
{E_0}^{g} = \tilde{M} + \mu \gamma
\end{equation}
where $\gamma =  \sqrt{ 1 - (Z\alpha)^2}$, and the form
\begin{equation}
  {\psi_0}^g(\vec p \, ) = \sqrt{1 \over 2 \tilde{M}} (\mu Z \alpha)^{-{3 \over 2}} 
  \left(
    \begin{array}{c}
         g(p) \chi_{-1 \mu}(\hat{p})  \\
        { \vec \sigma \cdot \vec p \over 2 \mu} f(p) \chi_{-1 \mu}(\hat{p})
    \end{array}
  \right),
\end{equation}
where $g(p)$ and $f(p)$ can be expressed in terms of the dimensionless
variable $q = p/(\mu Z \alpha)$,
\begin{eqnarray}
g(p) & = & {N \over 2q} {\rm sin}(\theta(1+\gamma)) [1 + q^2]^{-(1+\gamma)/2} \nonumber \\
f(p) & = & {N \over q^3(1+\gamma)} \left [ {{\rm sin} (\gamma \theta) \over \gamma} 
\sqrt{1+q^2} - q {\rm cos}(\theta(1+\gamma)) \right ] [1+q^2]^{-(1+\gamma)/2}.
\end{eqnarray}
Here $\theta \equiv {\rm tan}^{-1} q$ and the normalization factor $N$ is 
\begin{equation}
N = 2^{\gamma+3} \pi \Gamma(1+\gamma)\sqrt{(1+\gamma) \over
\Gamma(1+2\gamma)}.
\end{equation}
The $\chi_\kappa$'s are two-component eigenfunctions of $J^2$, $J_z$, $L^2$, and $S^2$
and are labeled by $\kappa = \mp (j+1/2)$ for $j=\ell \pm 1/2$ and $\mu$, the quantum number corresponding to $J_z$.  The spherical spinors are normalized so that $\chi^\dagger \chi$, when integrated over solid angle, gives one.
In the following we adopt the convention of working with Dirac wave functions with the
usual normalization, which we account for by multiplying a factor $1/(2\tilde{M})$ into
expressions for energy shifts, which always involve two Dirac wave functions.
We note that in most three-dimensional formalisms a lowest order potential that differs
from the one-Coulomb photon exchange kernel must be devised to obtain a Schr\"{o}dinger
or Dirac equation, and it in general has energy dependence, which leads to derivative
terms: neither complication appears in the present formalism.
With this definition of our lowest order problem, we now turn to the 
calculation of corrections to the lowest order energy, $\tilde{M} + {\cal E}$. 

\section{Perturbation expansion}

Perturbative corrections to the lowest order energy can be derived by calculating
the shift of the pole position in $\bar{G}$. To the order we are interested
in here, the shift can be shown to be
\begin{equation}
E-E_0 \equiv \Delta E_1 = {1 \over 2 \tilde{M}} \int {d^3 k   \, d^3 l \over (2\pi)^6} 
\bar{\psi}(\vec k \, ) [i\bar{K}(\vec k, \vec l;E) - iK_{1C}(\vec k, \vec l,E)] \psi(\vec l \, ).
\label{pert}
\end{equation}
Schematically we can write
\begin{equation}
\bar{K} = K_{1C} + K_{1T} + K_{CCX} + K_{CCs} + K_{1C}(S-S_0)K_{1C} + ...
\end{equation}
Here $1C$ and $1T$ refer to one Coulomb and one transverse photon exchange, $CCX$ is
the crossed ladder diagram with two Coulomb photons, $CCs$ is the seagull diagram with
two Coulomb photons, and the last term is the leading part of the correction induced in
the kernel by our change of propagators as given in Eq.~(\ref{Kbardef}). The 
term $K_{1C} S K_{1C}$ is an uncrossed ladder diagram, denoted $K_{CC}$, and it is easy to see 
that the term $-K_{1C} S_0 K_{1C}$ is equivalent to $-K_{1C}$ when used to evaluate $\Delta E_1$. 
The net effect, illustrated in Fig.~1,  is that
\begin{equation}
\bar{K} = K_{1T} + K_{CC} + K_{CCX} + K_{CCs} + ...
\end{equation}
Only these four terms need be considered to obtain the corrections of order 
$m^2 (Z\alpha)^4/M$, and we now turn to their evaluation.
\begin{figure}
\includegraphics[width=4.0in,trim=2.0in 7.0in 1.0in 1.6in]{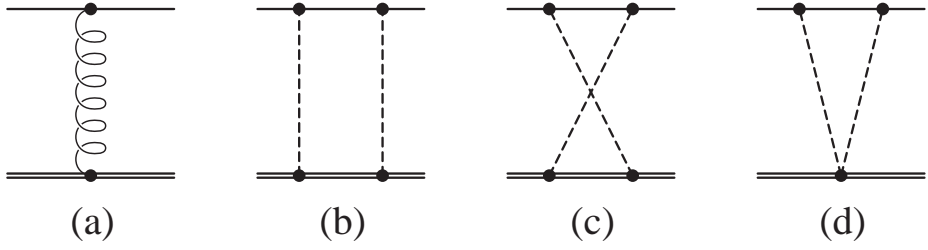}
\caption{Contributions to energy levels of hydrogenlike ions at order $m^2 (Z \alpha)^2/M$.  Initial and final wavefunctions are implicit.  Vertices on the top (electron) line are the usual spin-1/2 QED factors $-i e \gamma_\mu$ (where $e=\vert e \vert$).  Vertices on the bottom (nucleus) line are those appropriate for spin-0 QED: $i Z e (p+p')_\mu$ for the one-photon vertex and $2 i (Z e)^2 g_{\mu \nu}$ for the seagull vertex.  Part (a) represents the exchange of a transverse photon; (b) represents the exchange of two Coulomb ladder photons; (c) represents the Coulomb-Coulomb crossed ladder; and (d) represents the Coulomb-Coulomb seagull.}
\label{fig1}
\end{figure}

\subsection{One-transverse photon exchange}

The transverse photon propagator with momentum $q$ depends on both $q_0$ and $\vec q$.
It simplifies in our formalism, which forces $q_0=0$, and $K_{1T}$ gives the energy 
shift
\begin{equation} \label{integral_for_1T}
\Delta E_{1T} = -{4 \pi Z \alpha \over 2 \tilde{M}} \int {d^3 k  \, d^3 l \over (2 \pi)^6} 
{1 \over (\vec q \, ^2)^2} [ \vec q \, ^2 \psi^{\dagger}(\vec k \, ) \vec \alpha 
\cdot (\vec k + \vec l \, ) \psi(\vec l \, ) - \vec q \cdot (\vec k + \vec l \, ) 
\psi^{\dagger}(\vec k \, ) \vec \alpha \cdot \vec q \psi(\vec l \, )],
\end{equation}
where $\vec q = \vec k - \vec l$. If we approximate the Dirac wave functions in terms of
Schr\"{o}dinger wave functions through
\begin{equation}
\label{NRapprox}
  \psi(\vec p \, ) = 
  \left(
    \begin{array}{c}
         \phi_{NR}(p) \chi_{\kappa \mu}(\hat{p})  \\
         { \vec \sigma \cdot \vec p \over 2 \mu} \phi_{NR}(p) \chi_{\kappa \mu}(\hat{p})
    \end{array}
  \right),
\end{equation}
this simplifies to
\begin{eqnarray}
\Delta E_{1T}(NR) & = & - {4 \pi Z \alpha \over 4 m \tilde{M}} \int \frac{d^3 k}{(2 \pi)^3} 
\int \frac{d^3 l }{(2 \pi)^3}  \, 
\phi^\dagger_{NR}(k) \phi_{NR}(l) ~ \left [ {|\vec k + \vec l \, |^2 + 2i \vec \sigma \cdot 
(\vec k \times \vec l \, ) \over q^2} - { (k^2 -l^2)^2 \over q^4} \right ] \nonumber \\
& = & - {4 \pi Z \alpha \over m \tilde{M}}  \int \frac{d^3 k}{(2 \pi)^3} 
\int \frac{d^3 l }{(2 \pi)^3}  \, 
\phi^\dagger_{NR}(k) \phi_{NR}(l) ~ \left [ {k^2 l^2 - (\vec k \cdot \vec l \, )^2 \over q^4} + 
{i \sigma \cdot (\vec k \times \vec l \, ) \over 2q^2} \right ],
\end{eqnarray}
where the spherical spinors are understood.
This can be Fourier transformed into coordinate space, leading to spin independent and
spin dependent operators $H_R$ and $H_{SO}$,
\begin{eqnarray}
H_R &=& -\frac{Z \alpha}{2 \mu^2 r} \left ( \delta_{i j} + \hat x_i \hat x_j \right ) p_i p_j \cr
H_{SO} &=& \frac{Z \alpha}{4 \mu^2 r^3} \vec L \cdot \vec \sigma \, .
\end{eqnarray}
After working out the expectation value of $H_R$ one finds
\begin{equation}
\Delta E_{T} =  \frac{m}{M} m (Z \alpha)^4 \left \{ \frac{1}{n^4} + 
\frac{\delta_{\ell,0}}{n^3} - \frac{3}{n^3 (2 \ell+1)} \right \} + \frac{m}{M} <2 H_{SO} > \, ,
\end{equation}
where
\begin{equation}
<H_{SO}> = {\mu (Z\alpha)^4 (j(j+1)-\ell(\ell+1)-3/4) \over 2 n^3 \ell (\ell+1)(2\ell+1)}.
\end{equation}
It is of course straightforward to simply use exact wavefunctions and evaluate the integral (\ref{integral_for_1T})
numerically. The results of doing this for the ground state using the adaptive multidimensional 
integration program VEGAS \cite{VEGAS} are shown in Fig.~2, where the exact result is compared 
with the NR approximation. As is also typical for the nonrecoil case, significant differences
that would be poorly treated with an expansion in $Z\alpha$ arise at high $Z$. We note that a 
fit can be carried out, giving
\begin{equation}
\Delta E_{1T}(1s) = {m^2 (Z\alpha)^4 \over \tilde{M}} \left [ -1 - 1.50(1) (Z\alpha)^2 + ... \right ]
\end{equation}
consistent with the known $(Z\alpha)^6$ behavior.
\begin{figure}
\includegraphics[width=5.0in,trim=0.4in 0.0in 1.0in 1.0in]{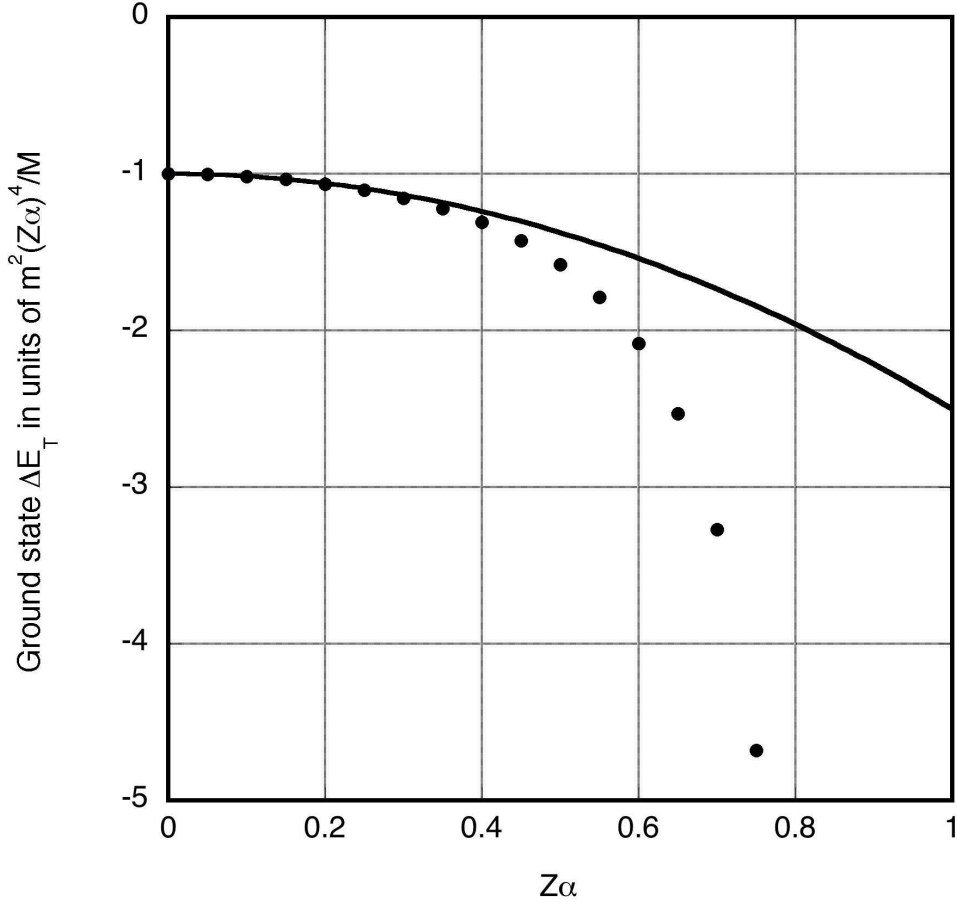}
\caption{Plot of $\Delta E_T$ for the ground state as a function of $Z \alpha$ in units of $m^2 (Z \alpha)^4/M$.  The data points represent the result of a numerical evaluation of (\ref{integral_for_1T}) without approximation.  The solid curve is the function $-1-(3/2)(Z \alpha)^2$, which includes the first relativistic correction.  We note that the expansion in $Z \alpha$ works well for small values of $Z$ but fails badly for large $Z$. }
\label{fig2}
\end{figure}

\subsection{Coulomb-Coulomb ladder}

The diagram which requires the greatest care is the two-Coulomb photon ladder diagram, as
it has a binding singularity. In addition, a new feature, present in a number of loop diagrams
when the nucleus is treated as a scalar, is poor convergence in the integration over the
fourth component of momentum, $q_0$, when Coulomb photons are present. Coulomb photon
propagators, being independent of $q_0$, provide no convergence, and the remaining $q_0$
is nominally logarithmically divergent, though that divergence vanishes by symmetry.
We regulate this near divergence by introducing a factor $\Lambda^2/(q_0^2 + \Lambda^2)$ 
and subsequently taking $\Lambda$ to infinity. This procedure introduces a term
that will be shown to cancel when 
a gauge invariant set of graphs is considered. Including this factor into the diagram of Fig.~1b
gives the energy shift
\begin{eqnarray}
\Delta E_{CC} =  i{(4\pi Z \alpha)^2 \over 2 \tilde{M}} \int {d^4 q \over (2\pi)^4} 
\int {d^3k  \, d^3l \over (2\pi)^6} {(2\tilde{M} - q_0)^2 \over (\tilde{M} - q_0)^2 - \vec q \, ^2 -M^2 +
i \delta}  \nonumber \\ {\Lambda^2 \over \Lambda^2 + q_0^2}
{1 \over |\vec k - \vec q \, |^2 |\vec q - \vec l \, |^2} {\bar{\psi}(\vec k \, ) \gamma_0[
({\cal E} + q_0)\gamma_0 - \vec \gamma \cdot \vec q + m]\gamma_0 \psi(\vec l \, ) \over
(({\cal E} + q_0)^2 - \vec q \, ^2 - m^2 + i \delta)}.
\end{eqnarray}
The Dirac equation can be used to carry out the $\vec k$ and $\vec l$ integrations,
leaving
\begin{eqnarray}
\Delta E_{CC} =   {i \over 2 \tilde{M}} \int {d^4 q \over (2\pi)^4} 
{(2\tilde{M} - q_0)^2 \over (\tilde{M} - q_0)^2 - \Omega_q^2 +
i \delta} {\Lambda^2 \over \Lambda^2 + q_0^2}  \nonumber \\
{\bar{\psi}(\vec q \, ) ({\cal E}\gamma_0 - \vec \gamma \cdot \vec q - \mu)
[ ({\cal E} + q_0)\gamma_0 - \vec \gamma \cdot \vec q + m] ({\cal E} \gamma_0 - 
\vec \gamma \cdot \vec q - \mu) \psi(\vec q \, ) \over
(({\cal E} + q_0)^2 - \omega_q^2 + i \delta)},
\end{eqnarray}
where $\Omega_q = \sqrt{\vec q \, ^2 + M^2}$ and $\omega_q = \sqrt{\vec q \, ^2 + m^2}$. We
note the `mismatch' in the numerator between terms with $\mu$ and with $m$: the former
come from the formalism, and the latter from the electron propagator in the diagram,
which is not altered by the choice of formalism. We now 
carry out the $q_0$ integration by closing above with Cauchy's theorem: once this is
done we are free to introduce a new four-vector $q = ( {\cal E}, \vec q \, )$. Three terms
result, with the simplest arising from the regulator,
\begin{equation}
\Delta E_{CC1} =   {1 \over 4 \tilde{M}} \int \frac{d^3 q}{(2 \pi)^3} 
\bar{\psi}(\vec q \, ) (\not\!q - \mu) \gamma_0  ( \not\!q - \mu) \psi(\vec q \, ).
\end{equation}
This term contributes in order $m^2(Z\alpha)^4/M$, but as it will be shown to cancel we
do not evaluate it explicitly.  The other two terms are both non-recoil, with the most
sensitive being
\begin{eqnarray}
\Delta E_{CC2} & = &   {1 \over 2\tilde{M}} \int \frac{d^3 q}{(2 \pi)^3} 
(\tilde{M} + \Omega_q)^2 {1 \over 2 \Omega_q}   \nonumber \\
&  & {\bar{\psi}(\vec q \, ) (\not\!q - \mu)
[ ({\cal E} + \tilde{M} - \Omega_q)\gamma_0 - \vec \gamma \cdot \vec q + m] 
(\not\!q - \mu) \psi(\vec q \, ) \over
({\cal E} + \tilde{M}-\Omega_q)^2 - \omega_q^2 }.
\end{eqnarray}
If we use the fact that
\begin{equation}
\frac{(\tilde M + \Omega_q)^2}{(2\tilde{M})(2 \Omega_q)} = 1 + O(1/M^2) \, 
\end{equation}
this simplifies to
\begin{eqnarray}
\Delta E_{CC2} =   \int \frac{d^3 q}{(2 \pi)^3} {\bar{\psi}(\vec q \, ) (\not\!q - \mu)
[({\cal E} + \tilde{M} - \Omega_q)\gamma_0 - \vec \gamma \cdot \vec q + m] 
(\not\!q  - \mu) \psi(\vec q \, ) \over
({\cal E} + \tilde{M}-\Omega_q)^2 - \omega_q^2  }.
\end{eqnarray}
We proceed by rearranging the interior numerator in the above as follows:
\begin{eqnarray} \label{CSC2num}
\gamma^0 ({\cal E} + \tilde M - \Omega_q) - \vec \gamma \cdot \vec q + m &=& 
(\not\!q + \mu)
+\gamma^0 (\tilde M-\Omega_q) + (m-\mu) \cr
&=& (\not\!q +\mu) + \frac{m}{M} \biggl \{ \gamma^0 \left ( m-{\vec q \, ^2 \over 2m} \right ) + m \biggr \} \cr
&=& (\not\!q+\mu) + \frac{m}{M} \biggl \{ \gamma^0 {\cal E} + 
\gamma^0 (m-{\cal E}) + m - \gamma^0 {\vec q \, ^2 \over 2m} \biggr \} \cr
&=& (\not\!q+\mu) + \frac{m}{M} \biggl \{ (\not\!q + \mu) + 
\gamma^0 (m-{\cal E}) \cr
&\hbox{}& + \vec \gamma \cdot \vec q - \gamma^0 {\vec q \, ^2 \over 2m}  \biggr \} \, ,
\end{eqnarray}
where in the last manipulation we have replaced $m$ with $\mu$, with the difference
being higher order in $1/M$. If we now define $\kappa = q^2 - \mu^2$ and restore the
factors $\not\!q - \mu$ on the left and right of the interior numerator we get
\begin{eqnarray} \label{CSC2nump}
\kappa \left (1 + {m \over M} \right ) (\not\!q - \mu) + {m \over M} (\not\!q - \mu) 
\left [ \gamma^0 (m-{\cal E}) + \vec \gamma \cdot \vec q - \gamma^0 {\vec q \, ^2 \over 2m} \right ]  
\, (\not\!q - \mu).
\end{eqnarray}
We further make the expansion of the denominator
\begin{eqnarray} \label{CSC2den}
\bigl ({\cal E} + \tilde M - \Omega_q \bigr )^2-\omega_q^2 &=& \bigl ( {\cal E}^2 - 
\vec q \, ^2 - \mu^2 \bigr ) + 2 {\cal E} (\tilde M-\Omega_q) - (m^2-\mu^2) + O(1/M^2) \cr
&=& \kappa + \frac{{\cal E}}{M} \bigl ( 2m^2-\vec q \, ^2) - 2m^3/M + O(1/M^2) \cr
&=& \kappa + \frac{1}{M} \left \{ {\cal E} ({\cal E}^2 - \vec q \, ^2 - m^2) - {\cal E}^3 + 
3 {\cal E}m^2-2m^3 \right \} + \cdots \cr
&=& \kappa + \frac{1}{M} \left \{ {\cal E} \kappa - (m-{\cal E})^2 (2m+{\cal E}) + 
\cdots \right \} \cr
&=& \kappa \left [ 1 +\frac{{\cal E}}{M} - \frac{(m-{\cal E})^2 (2m+{\cal E})}{M \kappa} + 
\cdots \right ] \, .
\end{eqnarray}
Combining these two forms then gives
\begin{eqnarray} \label{CSC2expanded}
\Delta E_{CC2} &=& \int \frac{d^3 q}{(2 \pi)^3} \bar \psi(\vec q \,) (\not\!q-\mu) 
\bigg \{ 1 + \frac{(m-{\cal E})}{M} + \frac{(m-{\cal E})^2 (2m+{\cal E})}{M \kappa} \cr
&\hbox{}& + \frac{m}{M \kappa} \left ( \gamma^0 (m-{\cal E}) + \vec \gamma \cdot \vec q - 
\gamma^0 \frac{\vec q \, ^2}{2m} \right ) (\not\!q-\mu) \bigg \} \psi(\vec q \,) \, .
\end{eqnarray}
If we label the terms in the curly brackets in (\ref{CSC2expanded}) parts 1-6,  the various 
contributions are
\begin{equation}
\Delta E_{CC21} = <V>_{\rm Dirac} \, ,
\end{equation}
\begin{equation}
\Delta E_{CC22} = \frac{m}{M} m (Z\alpha)^4 \biggl \{ \frac{-1}{2n^4} \biggr \} \, ,
\end{equation}
\begin{equation}
\Delta E_{CC23} = \frac{m}{M} m (Z\alpha)^4 \biggl \{ \frac{3}{8n^4} \biggr \} \, ,
\end{equation}
\begin{equation}
\Delta E_{CC24} = \frac{m}{M} m (Z\alpha)^4 \biggl \{ \frac{-1}{4n^4} \biggr \} \, ,
\end{equation}
\begin{equation}
\Delta E_{CC25} = \frac{m}{M} m (Z\alpha)^4 \biggl \{ \frac{-1}{2n^4} + \frac{2}{n^3 (2 \ell+1)} -
\frac{\delta_{\ell,0}}{n^3} \biggr \} - \frac{m}{M} <2 H_{SO} > \, ,
\end{equation}
\begin{equation}
\Delta E_{CC26} = \frac{m}{M} m (Z\alpha)^4 \biggl \{ \frac{-1}{4n^4} + \frac{1}{n^3 (2 \ell+1)}  
\biggr \} \, .
\end{equation}
To cancel the factor of $\kappa$ in the denominator, we note that its nonrelativistic limit
is $-\vec q \, ^2 - (\mu Z \alpha)^2/n^2$, so that the Schr\"{o}dinger equation reads
\begin{equation}
{- \kappa \over 2 \mu} \phi_{NR}(\vec q \, ) = 4 \pi Z\alpha \int \frac{d^3 p}{(2 \pi)^3} 
{1 \over |\vec q - \vec p \, |^2} \phi_{NR}(\vec p \, ).
\end{equation}
The strategy for the last three terms is then to `undo' the Dirac equation so that the Coulomb 
potential is explicitly present, take the nonrelativistic limit as in Eq.~(\ref{NRapprox}), and 
then use the Schr\"{o}dinger equation.  We illustrate this with $\Delta E_{CC25}$,
\begin{eqnarray} 
\Delta E_{CC25} &=& \int \frac{d^3 q}{(2 \pi)^3} \frac{m}{M \kappa} \bar \psi(\vec q \,) (\not\!q-\mu) 
\vec \gamma \cdot \vec q \, (\not\!q-\mu) \psi(\vec q \,) \, \nonumber \\
 &=& {(4 \pi Z \alpha)^2 \over (2 \pi)^9}  \int {d^3 k  \, d^3 q  \, d^3 l \over |\vec k - \vec q \, |^2
|\vec q - \vec l \, |^2} \frac{m}{M \kappa} 
\bar \psi(\vec k \,) \gamma_0 \vec \gamma \cdot \vec q  \gamma_0 \psi(\vec l \,) \, \nonumber \\
 & \approx & - {(4 \pi Z \alpha)^2 \over (2 \pi)^9} \int {d^3 k  \, d^3 q  \, d^3 l \over |\vec k - \vec q \, |^2
|\vec q - \vec l \, |^2} \frac{m}{2 M \kappa} 
{\phi^{\dagger}}_{NR}(\vec k \,) (\vec \sigma \cdot \vec q  \, \vec \sigma \cdot \vec l +
\vec \sigma \cdot \vec k  \, \vec \sigma \cdot \vec q \, ) \phi_{NR}(\vec l \,) \, \nonumber \\
 & \approx & {4 \pi Z \alpha \over (2 \pi)^6} \int {d^3 q  \, d^3 l \over 
|\vec q - \vec l \, |^2} \frac{1}{2 m M} 
{\phi^{\dagger}}_{NR}(\vec q \,) \left ( \vec q \cdot \vec l + i \vec \sigma \cdot (\vec q \times \vec l \, ) \right )
\phi_{NR}(\vec l \,).
\end{eqnarray}
The total contribution of term $CC2$ is then
\begin{equation}
\Delta E_{CC2} = <V>_{\rm Dirac} + \frac{m}{M} m (Z\alpha)^4 \biggl \{ \frac{-9}{8n^4} - 
\frac{\delta_{\ell,0}}{n^3} + \frac{3}{n^3 (2 \ell+1)}  \biggr \} - \frac{m}{M} <2 H_{SO} > \, ,
\end{equation}
where the formalism subtracts off the first, non-recoil term.  It is of interest that
had we used the formalism with ${\cal E} \rightarrow \epsilon$, $\tilde{M} \rightarrow M$
mentioned above, the cancellation, while still removing the nonrecoil term, would leave
a contribution that can be shown to start with the term ${m^2 (Z\alpha)^2 \over 2Mn^2}$, the standard
reduced mass contribution to the nonrelativistic energy. As we have chosen to build this
into our lowest order solution, the cancellation is finer, and leaves terms starting in
order $m^2(Z\alpha)^4/M$. 

The remaining part of the CC calculation involves closing around a negative energy electron
pole, which while leading to higher powers of $Z\alpha$ than when the nuclear pole is
encircled, is also nonrecoil. Its full contribution is
\begin{equation}
\Delta E_{CC3} = {1 \over 2\tilde{M}} \int {d^3 q  \over (2 \pi)^3 2 \omega_q}
(2\tilde{M} + {\cal E} + \omega_q)^2 
{\bar{\psi}(\vec q \, ) (\not\!q - \mu)
[  -\omega_q\gamma_0 - \vec \gamma \cdot \vec q + m] 
(\not\!q - \mu) \psi(\vec q \, ) \over
({\cal E} + \tilde{M}+\omega_q)^2 - \Omega_q^2 },
\end{equation}
but we can approximate ${\cal E}=m$, $\omega_q = m$, and $\Omega_q=M$ in the nuclear propagator, 
leading to the simpler expression
\begin{equation}
\Delta E_{CC3}  =  {1 \over 2\tilde{M}} \int {d^3 q  \over (2 \pi)^3 2 \omega_q}
\frac{(2\tilde{M} + 2m )^2}{ (\tilde{M}+ 2m)^2 - M^2 }
\bar{\psi}(\vec q \, ) (\not\!q - \mu) [  -\omega_q\gamma_0 - \vec \gamma \cdot \vec q + m] 
(\not\!q - \mu) \psi(\vec q \, ) \, .
\end{equation}
We will show below that although this term is nonrecoil, starting in order $m (Z\alpha)^4$,
the nonrecoil part cancels with a contribution from the crossed Coulomb ladder.

\subsection{Crossed Coulomb diagram}

The crossed ladder (CCX) diagram of Fig.~1c is given by
\begin{eqnarray}
\Delta E_{CCX} =  i{(4\pi Z \alpha)^2 \over 2 \tilde{M}} \int {d^4 q \over (2\pi)^4} 
\int {d^3k  \, d^3l \over (2\pi)^6} {(2\tilde{M} + q_0)^2 \over 
(\tilde{M} + q_0)^2 - |\vec q - \vec k - \vec l \, |^2 -M^2 + i \delta}  \nonumber \\
{\Lambda^2 \over q_0^2 + \Lambda^2} {1 \over |\vec k - \vec q \, |^2 |\vec q - \vec l \, |^2} 
{\bar{\psi}(\vec k \, ) \gamma_0[ ({\cal E} + q_0)\gamma_0 - \vec \gamma \cdot \vec q + m]\gamma_0 
\psi(\vec l \, ) \over (({\cal E} + q_0)^2 - \omega_q^2 + i \delta)}.
\end{eqnarray}
Taking a pole of the regulator term gives the same result as with the ladder,
thus doubling $\Delta E_{CC1}$. While the
pole from the nuclear line enters in the entirely negligible order $m^5/M^4$, the
electron pole contributes at the level of non-recoil fine structure, and is
\begin{eqnarray}
\Delta E_{CCX} =  {(4 \pi Z \alpha)^2 \over  2 \tilde{M}} \int \frac{d^3 q}{(2 \pi)^3} 
\int {d^3k  \, d^3l \over (2\pi)^6} {(2\tilde{M} - {\cal E} - \omega_q)^2 \over 
(\tilde{M} - {\cal E} - \omega_q)^2 - |\vec q - \vec k - \vec l \, |^2 -M^2 }  \nonumber \\
{1 \over 2 \omega_q} {1 \over |\vec k - \vec q \, |^2 |\vec q - \vec l \, |^2} 
\bar{\psi}(\vec k \, ) \gamma_0[ - \omega_q \gamma_0 - \vec \gamma \cdot \vec q + m]\gamma_0 
\psi(\vec l \, ).
\end{eqnarray}
It is again legitimate to make the approximations ${\cal E} = m$, $\omega_q = m$,
and $M^2 +  |\vec q - \vec k - \vec l \, |^2 = M^2$, and further use of
the Dirac equation gives
the approximation
\begin{equation}
\Delta E_{CCX} =  {1 \over  2 \tilde{M}} \int {d^3 q \over (2 \pi)^3 2 \omega_q}
{(2\tilde{M} - 2m)^2 \over (\tilde{M} - 2m)^2 - M^2}  
\bar{\psi}(\vec q \, ) (\not\!q -\mu) [ - \omega_q \gamma_0 - \vec \gamma \cdot \vec q + m]
(\not\!q - \mu) \psi(\vec q \, ).
\end{equation}
This term can now be combined with the CC3 contribution, and the nonrecoil term can
easily be seen to cancel. There remains a nonvanishing recoil term of order $m^2(Z\alpha)^5/M$,
\begin{equation}
\Delta E_{CC3}+\Delta E_{CCX} =  {1 \over  2 \tilde{M}} \int {d^3 q \over (2 \pi)^3 2 \omega_q}
\bar{\psi}(\vec q \, ) (\not\!q -\mu) [ - \omega_q \gamma_0 - \vec \gamma \cdot \vec q + m]
(\not\!q - \mu) \psi(\vec q \, ),
\end{equation}
which we keep, although beyond the order of interest we are considering here, as it
has an interesting connection with the seagull diagram.

\subsection{Seagull diagram}
A novel feature of our formalism is the presence of so-called seagull graphs.
In Coulomb gauge the seagull graph consists of a Coulomb-Coulomb (CC) term and
a transverse-transverse term (TT), with the latter beyond our present order of interest.
Again regularizing the $q_0$ integration, the CC seagull graph (see Fig.~1d) contributes
\begin{eqnarray}
\Delta E_{CCs} & = & -2i{(4\pi Z \alpha)^2 \over 2 \tilde{M}} \int {d^4 q \over (2\pi)^4} 
\int {d^3k  \, d^3l \over (2\pi)^6} 
{\Lambda^2 \over \Lambda^2 + q_0^2} \nonumber \\
& & {1 \over |\vec k - \vec q \, |^2 |\vec q - \vec l \, |^2} {\bar{\psi}(\vec k \, ) \gamma_0[
({\cal E} + q_0)\gamma_0 - \vec \gamma \cdot \vec q + m]\gamma_0 \psi(\vec l \, ) \over
(({\cal E} + q_0)^2 - \vec q \, ^2 - m^2 + i \delta)}.
\end{eqnarray}
If we again close above to carry out the $q_0$ integration the regulator term contributes
$-2 \Delta E_{CC1}$: as $\Delta E_{CC1}$ was doubled from the crossed Coulomb diagram,
this completes the cancellation of contributions arising from the regulator term.
The other pole picks up a negative energy electron contribution, and gives,
using the Dirac equation,
\begin{equation}
\Delta E_{CCs} =  -2 {1 \over 2 \tilde{M}} \int {d^3 q \over (2 \pi)^3 2 \omega_q}
\bar{\psi}(\vec q \, )(\not\!q - \mu) [ -\omega_q \gamma_0 - \vec \gamma \cdot \vec q + m]
(\not\!q - \mu) \psi(\vec q \, ).
\end{equation}
The net result is that the only role of the  seagull diagram to order $m^2(Z\alpha)^4/M$ is 
in canceling the regulator terms from the ladder and crossed ladder, and in addition it
combines with $m^2(Z\alpha)^5/M$ terms coming from the negative energy pole terms
in those diagrams.

\section{Total at order $(Z \alpha)^4$}

The combination of the CC, CCX, CCs, T, and -C graphs gives
\begin{equation}
\Delta E = \frac{m}{M} m (Z \alpha)^4 \left \{ \frac{-1}{8n^4} \right \} \, .
\end{equation}
In order to find the total recoil contribution at this order, one must combine this with the 
recoil contribution from $\mu (f(n,j)-1)$, which is $-(m^2/M)(f(n,j)-1)$ where 
$f(n,j)=\epsilon/m$ is defined through Eq.~(\ref{reference_energies}).  The total recoil 
contribution through terms of order $(Z \alpha)^4$ is
\begin{eqnarray}
\Delta E_{\rm recoil} &=& \frac{m^2}{M} \biggl \{ - \bigl (f(n,j)-1\bigr ) - 
\frac{(Z \alpha)^4}{8n^4} \biggr \} \cr
&=& \frac{m^2}{M} \biggl \{ \frac{(Z \alpha)^2}{2n^2} + (Z \alpha)^4 \biggl 
( \frac{-1}{2n^4} + \frac{1}{2 n^3 (j+1/2)} \biggr ) \biggr \} \, .
\end{eqnarray}
This is the known Barker and Glover result \cite{Barker} for the recoil contribution at this order.

\section{Salpeter correction}

One of the first accomplishments of a fully relativistic treatment of the two-body bound state
problem was Salpeter's discovery \cite{Salpeter} that corrections of order $\alpha m/M$ 
times fine structure were present in hydrogenlike atoms: use of the older formalism of
the Breit equation had found no such contributions \cite{Arfken}. While we are not calculating 
all such terms here, we show how they arise from one and higher Coulomb exchanges that are
crossed by a transverse photon, illustrating with the graph of Fig.~3a. This gives rise to the
somewhat complicated expression
\begin{eqnarray}
\Delta E & = &  -{(4\pi Z \alpha)^3 \over 2 M} \int {d^4 k \over (2\pi)^4} 
\int {d^4 l \over (2\pi)^4} \int {d^3 p  \, d^3p' \over (2\pi)^6} {(2 M + 2 k_0-l_0) \over 
(M+k_0)^2-|\vec k - \vec p \, |^2 - M^2 + i \delta} \nonumber \\
& & {(2 M + k_0 -l_0) \over [(M + k_0 -l_0)^2 - |\vec k - \vec l \, |^2 - M^2 + i \delta]}  
{(k-2p)_j (\delta_{ij} - {k_i k_j \over \vec k^2}) \over k_0^2 - \vec k^2 + i \delta} 
{1 \over |\vec l - \vec p \, |^2}  {1 \over |\vec k - \vec l + \vec p\,'|^2} \nonumber \\
& & \bar{\psi}(\vec p\,') \gamma_i { 1 \over
({\cal E} + k_0)\gamma_0 - \vec \gamma \cdot (\vec k + \vec p \, ') - m+ i \delta} \gamma_0
{1 \over ({\cal E} + l_0)\gamma_0 - \vec \gamma \cdot \vec l - m + i \delta} \gamma_0
\psi(\vec p \, ).
\end{eqnarray}
\begin{figure}
\includegraphics[width=5in,trim=0.8in 7.0in 1.2in 1.5in]{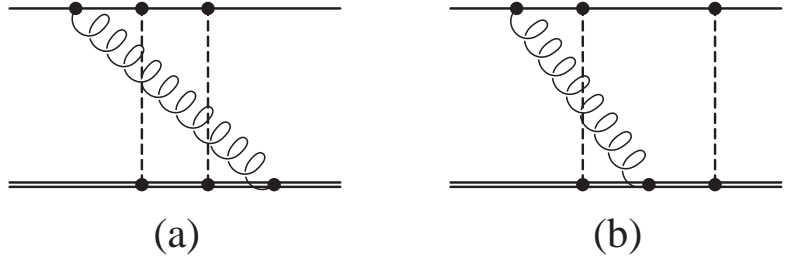}
\caption{Graphs contributing to the Salpeter correction of order $m^2 (Z \alpha)^5/M$.  In
graph (a) a transverse photon crosses two Coulomb photons.  In graph (b) the transverse
photon crosses a single Coulomb photon, with a Coulomb ladder photon on the side.  There
are two graphs like (b) since the ladder photon can be on either side.}
\label{fig3}
\end{figure}
We wish to show that this expression is dominated by a term that contains part of
the Dirac-Coulomb propagator of Eq.~(\ref{DiracCoulombprop}). Specifically, if we expand the momentum
space form of that equation in powers of the Coulomb potential, the term involving one 
potential is 
\begin{equation}
S_F^{1C}(\vec p, \vec p \, '; E) = { 1 \over E \gamma_0 - \vec \gamma \cdot \vec p\,' - m}
{ -4 \pi Z \alpha \over |\vec p\,' - \vec p \, |^2}
\gamma_0 {1 \over E \gamma_0 - \vec \gamma \cdot \vec p -m} ,
\label{1Cexp}
\end{equation}
where the distinction between $m$ and $\mu$ is dropped in this section as the graph being 
considered has a factor $m/M$. To see how this arises from the graph of Fig. 3a,
we first close the $k_0$ contour above and take the transverse photon pole to get
\begin{eqnarray}
\Delta E & = & i {(4\pi Z \alpha)^3 \over 4 M} \int {d^3 k \over (2\pi)^3} {1 \over k} 
\int {d^4 l \over (2\pi)^4} \int {d^3 p  \, d^3p' \over (2\pi)^6} {(2 M - 2 k-l_0) \over 
(M-k)^2-|\vec k - \vec p \, |^2 - M^2 } \nonumber \\
& & {(2 M - k -l_0) \over [(M - k -l_0)^2 - |\vec k - \vec l \, |^2 - M^2 + i \delta]}  
(-2p)_j \left (\delta_{ij} - {k_i k_j \over \vec k^2} \right ) 
{1 \over |\vec l - \vec p \, |^2}  {1 \over |\vec k - \vec l + \vec p\,'|^2} \nonumber \\
& & \bar{\psi}(\vec p\,') \gamma_i { 1 \over
({\cal E} - k)\gamma_0 - \vec \gamma \cdot (\vec k + \vec p \, ') - m} \gamma_0
{1 \over ({\cal E} + l_0)\gamma_0 - \vec \gamma \cdot \vec l - m + i \delta} \gamma_0
\psi(\vec p \, ).
\end{eqnarray}
Because we are dropping terms of order $m^3/M^2$, the first nuclear denominator simplifies 
to $-2Mk$, and further carrying out the $l_0$ integration by closing above gives a term from the
second nuclear denominator that forces $l_0 = -k$, giving
\begin{eqnarray}
\Delta E & = & - {(4\pi Z \alpha)^3 \over 4 M} \int {d^3 k \over (2\pi)^3} {1 \over k^2} 
\int {d^3 l \over (2\pi)^3} \int {d^3 p  \, d^3p' \over (2\pi)^6} 
(-2p)_j \left ( \delta_{ij} - {k_i k_j \over \vec k^2} \right ) 
{1 \over |\vec l - \vec p \, |^2}  {1 \over |\vec k - \vec l + \vec p\,'|^2} \nonumber \\
& & \bar{\psi}(\vec p\,') \gamma_i { 1 \over
({\cal E} - k)\gamma_0 - \vec \gamma \cdot (\vec k + \vec p \, ') - m} \gamma_0
{1 \over ({\cal E} - k)\gamma_0 - \vec \gamma \cdot \vec l - m } \gamma_0
\psi(\vec p \, ).
\end{eqnarray}
The Salpeter correction is associated with the region of integration in which
$k \sim m (Z\alpha)^2$ rather than the normal $m (Z\alpha)$. In this region one
can approximate $\vec k + \vec p\,' = \vec p\,'$ and $\vec k - \vec l + \vec p\,' =
- \vec l + \vec p\,'$ in the above. Using Eq.~(\ref{1Cexp}) then allows us to write
\begin{eqnarray}
\Delta E & = & \frac{(4\pi Z \alpha)^2}{4M}  \int {d^3 k \over (2\pi)^3} {1 \over k^2} 
\int {d^3 l \over (2\pi)^3} \int {d^3 p  \, d^3p' \over (2\pi)^6} 
(-2p)_j \left ( \delta_{ij} - {k_i k_j \over \vec k^2} \right ) 
{1 \over |\vec l - \vec p \, |^2}  \nonumber \\
& & \bar{\psi}(\vec p\,') \gamma_i S_F^{1C}(\vec p \, ', \vec l; {\cal E} -k) \gamma_0 \psi(\vec p \, ).
\end{eqnarray}
The same kinds of argument apply for any number of Coulomb exchanges, allowing the
replacement of $S_F^{1C}$ with $S_F$. Care is required, however, for the first term of 
the expansion of $S_F$, which requires an ultraviolet cutoff in the $k$ integration  
because of the approximations we have made. However, this term is finite when simply 
treated as a one loop diagram.  The actual calculation of the Salpeter correction 
would involve evaluating the one loop diagram without approximation and then evaluating 
the above expression and higher Coulomb exchanges by using $S_F-S_0$, a technique that 
is standard in self-energy calculations. Replacing $S_F^{1C}$ in the above with the
spectral representation of $S_F$ then gives
\begin{eqnarray}
\Delta E & = & \frac{(4\pi Z \alpha)^2}{4M}  \int {d^3 k \over (2\pi)^3} {1 \over k^2} 
\int {d^3 l \over (2\pi)^3} \int {d^3 p  \, d^3p' \over (2\pi)^6} 
(-2p)_j  \left ( \delta_{ij} - {k_i k_j \over \vec k^2} \right ) 
{1 \over |\vec l - \vec p \, |^2}  \nonumber \\
& & \sum_m {\bar{\psi}(\vec p\,') \gamma_i \psi_m(\vec p \, ')
\bar{\psi}_m(\vec l \,) \gamma_0 \psi(\vec p \, ) \over {\cal E} - E_m -k}.
\label{salpeterX1}
\end{eqnarray}
Replacing $-2p_j$ with $-2(p-l)_j$ leads to the integral
\begin{eqnarray}
& &\int {d^3 p \, d^3l \over (2\pi)^6} (p-l)_j  {{4 \pi Z \alpha} \over |\vec l - \vec p \, |^2} 
\bar{\psi}_m(\vec l \, ) \gamma_0 \psi(\vec p \, ) \nonumber \\
& = & -i Z \alpha \int d^3 x {x_j \over x^3} \bar{\psi}_m(\vec x \, ) \gamma_0 \psi(\vec x \, )  \nonumber \\
& = & ({\cal E} - E_m) < m | p_j | 0>
\label{salpeterX2}
\end{eqnarray}
where $\cal E$ is the Dirac energy of the state $\psi$ of interest, here taken to be the
ground state.  Eqs.~(\ref{salpeterX1}) and (\ref{salpeterX2}) lead to the relativistic 
generalization of Salpeter's \cite{Salpeter} Eq.~(45).  The replacement of $-2p_j$ by
$-2(p-l)_j$ arises from consideration of the reducible graph shown in Fig.~3b,
which comes from the formalism when three-photon exchange is considered, as described
in Ref.~\cite{poshfs}. In this diagram there are two nuclear propagators that depend
on $l_0$, with one of them leading to the replacement mentioned above, and the other
to a bound state singularity canceled by the formalism.

\section{NRQED calculation of the energy shift to order $(Z \alpha)^4$ including recoil}

We base our expression of NRQED on the work of Kinoshita and Nio \cite{Kinoshita96}.  
The NRQED Lagrangian for a particle of spin-1/2 and one of spin-0 has the form
\begin{eqnarray}
{\cal L} &=& \psi^\dagger \Biggl \{ i D_t + \frac{\vec D^2}{2m} + \frac{\vec D^4}{8 m^3}
+ c_F \frac{e \vec \sigma \cdot \vec B}{2m} \cr
&\hbox{}& + c_D \frac{e(\vec D \cdot \vec E - \vec E \cdot \vec D)}{8m^2}
+ c_S \frac{i e \vec \sigma \cdot (\vec D \times \vec E - \vec E \times \vec D)}{8m^2}
+ \cdots \Biggr \} \psi \cr
&\hbox{}& + \phi \Biggl \{  i D_t + \frac{\vec D^2}{2M} + \cdots \Biggr \} \phi + {\cal L}_{EM} \, ,
\end{eqnarray}
where $D_t = \partial_t + i e A^0$ and $\vec D = \vec \partial - i e \vec A$ for the electron, 
and similarly but with $e \rightarrow -Z e$ for the nucleus.  Electromagnetism is described 
by the usual Lagrangian ${\cal L}_{EM} = (-1/4) F_{\mu \nu} F^{\mu \nu}$, and we use Coulomb gauge 
in our calculations.  In terms of Feynman rules, we build on those given by Kinoshita and Nio 
in their Fig.~3.  The new rules include a Coulomb vertex for the nucleus $-Z e$, a dipole 
vertex for the nucleus $Z e (\vec p\,'+\vec p\,)/(2 M)$, and a propagator for the 
nucleus $(E-\vec p\,^2/(2M)+i \epsilon)^{-1}$.  The rule is to multiply all propagators 
by $i$, all vertices by $-i$, and to include an overall factor of $i$ when calculating energy 
shifts.  Loop integrals are done over all momenta with measure $d^4 k/(2 \pi)^4$.  The 
Bethe-Salpeter equation for NRQED (with lowest order propagators and vertices) is exactly 
the Schr\"odinger-Coulomb equation with reduced mass, which is also described in Kinoshita 
and Nio.  We use the symbol $\Psi(p)$ for the NRQED Bethe-Salpeter wave function.  For example, 
the ground state wave function is
\begin{equation}
\Psi(p) = (2 \pi) \delta(p_0) \psi(\vec p \,) \, ,
\end{equation}
where
\begin{equation}
\psi(\vec p\,) = \frac{16 \pi \xi^{5/2}}{(\vec p\,^2+\xi^2)^2} \chi_{-1 \mu}
\end{equation}
with $\xi = \mu Z \alpha$ and, for example, 
${\chi^{\dagger}}_{-1,\frac{1}{2}}=\frac{1}{\sqrt{4 \pi}} (1,0)$.

\begin{figure}
\includegraphics[width=5in,trim=1.0in 7.0in 1.0in 1.5in]{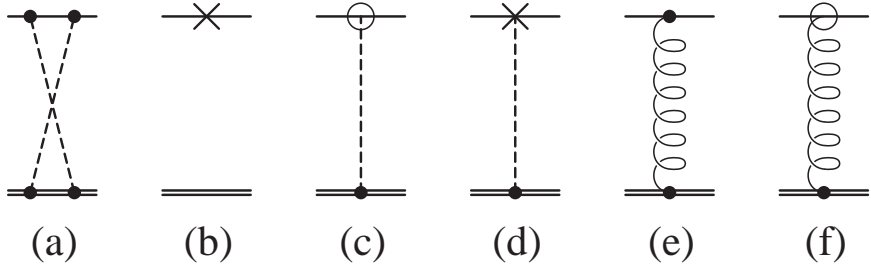}
\caption{NRQED contributions to energies at order $m^2 (Z \alpha)^4/M$.  Graph 
(a) is the two-Coulomb crossed ladder, (b) represents the $p^4$ relativistic kinetic energy 
correction to the electron line, (c) is the spin-orbit correction to the electron line (with Coulomb 
photon exchange), (d) is the Darwin correction to the electron line (with Coulomb photon exchange), 
(e) shows transverse photon exchange with dipole vertices on both electron and nuclear lines, and 
(f) represents transverse photon exchange with a Fermi vertex on the electron line and a dipole vertex on the nuclear line. }
\label{fig4}
\end{figure}
The relevant graphs for the calculation of energies up to $O((Z \alpha)^4)$ and including 
recoil (to first order: $m/M$) are shown in Fig.~\ref{fig4}.  In the calculation of bound state NRQED graphs we take the 
electron line to enter with momentum $({\cal E}_0, \vec p\,)$ and the nucleus with 
momentum $(0,-\vec p\,)$ where ${\cal E}_0 = -\mu (Z \alpha)^2/(2 n^2)$ is the Bohr energy level.

The crossed Coulomb ladder (Fig.~4a) has the form
\begin{eqnarray}
\Delta E_{CCX} &=& i \int \frac{d^3 q}{(2 \pi)^3} \frac{d^4 l}{(2 \pi)^4} \frac{d^3 p}{(2 \pi)^3}
\psi^\dagger(\vec q \,) (-ie) 
\frac{i}{l_0+{\cal E}_0-\frac{\vec l \,^2}{2m} + i \epsilon} (-ie) \psi(\vec p\,) \cr
&\hbox{}& \times \frac{i}{(\vec q - \vec l \,)^2}  \frac{i}{(\vec l - \vec p\,)^2} 
(i Z e) \frac{i}{l_0- \frac{(\vec l -\vec p -\vec q\,)^2}{2M} + i \epsilon} (i Z e) \, .
\end{eqnarray}
The poles of the $l_0$ integral
\begin{equation}
\int \frac{d l_0}{2 \pi i} \frac{1}{l_0+{\cal E}_0 - \frac{\vec l\,^2}{2m} + 
i \epsilon} \frac{1}{l_0- \frac{(\vec l-\vec p -\vec q\,)^2}{2m} + i \epsilon} = 0.
\end{equation}
are both on the same side of the real axis.  It follows that the $l_0$ integral vanishes, as does
the crossed Coulomb ladder contribution: $\Delta E_{CCX}=0$.

The relativistic kinetic energy correction (Fig.~4b) is
\begin{eqnarray}
\Delta E_K &=& i \int \frac{d^3 q \, d^3 p}{(2 \pi)^6} \, \psi^\dagger(\vec q\,) (-i) \left 
( \frac{-\vec p\,^4}{8m^3} \right )
(2 \pi)^3 \delta(\vec p - \vec q\,) \psi(\vec p\,) \cr
&=& \left ( \frac{\mu}{m} \right )^3 \left < H_K \right > \cr
&\approx& \left ( 1 - \frac{3m}{M} \right ) \left < H_K \right > \, ,
\end{eqnarray}
where
\begin{equation}
\left < H_K \right > = \int \frac{d^3 p}{(2 \pi)^3}  \psi^\dagger(\vec p\,) \left ( \frac{-\vec p\,^4}{8\mu^3} 
\right ) \psi(\vec p\,)
=\mu (Z \alpha)^4 \left ( \frac{3}{8 n^4} - \frac{1}{(2 \ell+1)n^3} \right ) \, .
\end{equation}

The spin-orbit correction to the electron line (Fig.~4c) is
\begin{eqnarray}
\Delta E_{SO} &=& i \int \frac{d^3 q \, d^3 p}{(2 \pi)^6} \, \psi^\dagger(\vec q\,) (-i) \frac{ie}{4m^2} 
\left ( \vec q \times \vec p  \, \right ) \cdot \vec \sigma  \psi(\vec p\,) \frac{i}{\vec k\,^2} 
(i Z e) \cr 
&=& \frac{i}{4m^2} \int \frac{d^3 q \, d^3 p}{(2 \pi)^6} \, \psi^\dagger(\vec q\,) \left ( \vec q 
\times \vec p \cdot \vec \sigma \right ) \psi(\vec p\,) V_C(\vec k\,) \cr
&=&  \left ( \frac{\mu}{m} \right )^2 \left < H_{SO} \right > \cr
&\approx& \left ( 1 - \frac{2m}{M} \right ) \left < H_{SO} \right > \, ,
\end{eqnarray}
where $\vec k = \vec q - \vec p$ and
\begin{equation}
\left < H_{SO} \right > = \left < \frac{Z \alpha}{4 \mu^2 r^3} \vec L \cdot \vec \sigma \right >
\end{equation}
as before.

The Darwin correction to the electron line (Fig.~4d) is
\begin{eqnarray}
\Delta E_D &=& i \int \frac{d^3 q \, d^3 p}{(2 \pi)^6} \, \psi^\dagger(\vec q\,) \frac{ie}{8m^2} 
\left | \vec q - \vec p \, \right |^2   \psi(\vec p\,) \frac{i}{\vec k\,^2} (i Z e) \cr
&=& \left ( \frac{\mu}{m} \right )^2 \frac{4 \pi Z \alpha}{8 \mu^2} \left | \psi(0) \right |^2 \cr
&\approx& \left ( 1 - \frac{2m}{M} \right ) \left < H_{D} \right > \, ,
\end{eqnarray}
where 
\begin{equation}
\left < H_D \right > = \frac{4 \pi Z \alpha}{8 \mu^2} \left | \psi(0) \right |^2 = \mu (Z \alpha)^4
\frac{\delta_{\ell,0}}{2n^3} \, .
\end{equation}

The dipole-dipole transverse photon exchange contribution (Fig.~4e) is
\begin{eqnarray}
\Delta E_{DD} &=& i \int \frac{d^3 q \, d^3 p}{(2 \pi)^6} \, \psi^\dagger(\vec q\,) \frac{ie}{2m} (p+q)^i
\frac{i \delta^T_{i j}(\vec k\,)}{-\vec k\,^2} \psi(\vec p\,) \frac{-i Z e}{2M} (-p-q)^j \cr
&=& \frac{-4 \pi Z \alpha}{m M} \int \frac{d^3 q \, d^3 p}{(2 \pi)^6} \, \psi^\dagger(\vec q\,)
\frac{\vec p\,^2 \vec q\,^2 - (\vec p \cdot \vec q\,)^2}{\vec k\,^4} \psi(\vec p\,) \cr
&=& \frac{m}{M} \left < H_R \right > \, ,
\end{eqnarray}
just as in the relativistic calculation of one transverse photon exchange.  We note that the 
expectation value $\left < H_R \right >$ can be written as
\begin{eqnarray}
\left < H_R \right > &=& \left \{ \frac{1}{n^4} + \frac{\delta_{\ell,0}}{n^3} - 
\frac{3}{n^3 (2 \ell+1)} \right \}
\mu (Z \alpha)^4 \cr
&=& 3 \left < H_K \right > + 2 \left < H_D \right > - \frac{1}{8 n^4} \mu (Z \alpha)^4 \, .
\end{eqnarray}

Finally, the Fermi correction (Fig.~4f) is
\begin{eqnarray}
\Delta E_{F} &=& \frac{-4 \pi Z \alpha}{4 m M} \int \frac{d^3 q \, d^3 p}{(2 \pi)^6} \,  \psi^\dagger(\vec q\,)
\frac{-i \epsilon_{i j k} (q-p)_j \sigma_k \delta^T_{i n}(\vec k\,) (p+q)_n}{\vec k\,^2} 
\psi(\vec p\,) \cr
&=& \frac{2i}{4 m M}  \int \frac{d^3 q \, d^3 p}{(2 \pi)^6} \,  \psi^\dagger(\vec q\,) \vec q \times \vec p 
\cdot \vec \sigma \psi(\vec p\,) V_C(\vec k\,) \cr
&=& \frac{m}{M} \left < \frac{Z \alpha}{2 m^2 r^3} \vec L \cdot \vec \sigma \right > \cr
&\approx& \frac{m}{M} \left < 2 H_{SO} \right > \, .
\end{eqnarray}

The sum of all contributions is
\begin{eqnarray}
\Delta E &=& \Delta E_{CCX}+\Delta E_K+\Delta E_{SO}+\Delta E_D
+\Delta E_{DD}+\Delta E_F \cr
&=& \left ( 1 - \frac{3m}{M} \right ) 
\left < H_K \right >
+ \left ( 1 - \frac{2m}{M} \right ) \left < H_{SO} \right >
+ \left ( 1 - \frac{2m}{M} \right ) \left < H_{D} \right > \cr
&\hbox{}&
+ \frac{m}{M} \left ( 3 \left < H_K \right > + 2 \left < H_D \right > - 
\frac{1}{8 n^4} \mu (Z \alpha)^4 \right )
+ \frac{m}{M} \left < 2 H_{SO} \right > \cr
&=& \left < H_K + H_{SO} + H_D \right > - \frac{m^2}{M} \frac{(Z \alpha)^4}{8n^4} \, ,
\end{eqnarray}
which again is the known Barker and Glover result for the fine structure with recoil correction.
 
\section{Conclusions}

A form of the Bethe-Salpeter equation of particular simplicity has been introduced that
can be applied to the entire hydrogen isoelectronic sequence. We have shown that the
power series expansion of the 1T kernel is nonperturbative at high $Z$, demonstrating
the need for a complete numerical calculation for all kernels. Such a calculation has 
been done using a Green's function formalism by Shabaev and collaborators
\cite{Victor}, \cite{Victors}, but it is always desirable in QED to have checks on these 
complex calculations. We are presently calculating the remaining one-loop diagrams
that enter in order $m^2/M (Z\alpha)^5$. As an indication of the numerical importance
of these calculations for the transition discussed in the introduction, we note that
Ref. \cite{Victors} finds a correction of -0.04 eV for the $2p_{1/2}-2s$ transition in 
hydrogenic uranium, to be compared to the 0.207 eV discrepancy presumably dominated
by the two-loop Lamb shift. However, this is only part of the effect of recoil for
lithiumlike uranium, and the question of relativistic corrections to mass polarization
cannot be addressed in our formalism, which is strictly a two-body approach. We are
presently investigating the relatively unexplored problem of forming many-particle
generalizations of the Bethe-Salpeter equation that have the three-dimensional and 
relativistic aspects of the equation described in the present work.

\acknowledgements

The work of J.S. was supported in part by NSF Grant No. PHY-0451842, and
grateful thanks are given to K.T. Cheng and the hospitality of LLNL, where this work was
started. Useful conversations with S. Morrison are also acknowledged.

\end{document}